\documentclass[twocolumn,showpacs,preprintnumbers,amsmath,amssymb]{revtex4}

\usepackage{graphicx}
\usepackage{dcolumn}
\usepackage{bm}

\begin{document}
\preprint{APS/123-QED}
\title{Swing wave-wave interaction: Coupling between fast magnetosonic and Alfv{\'e}n waves}

\author{T.V. Zaqarashvili}
 \altaffiliation[Also at ]{Abastumani Astrophysical Observatory, Al.
Kazbegi ave. 2a, 380060 Tbilisi, Georgia}
\author{B. Roberts}
\affiliation{School of Mathematics and Statistics,
University of St
Andrews, St Andrews, Fife KY16 9SS, Scotland
}%

\date{\today}

\begin{abstract}
 We suggest a mechanism of energy transformation from fast magnetosonic waves propagating
across a magnetic field to Alfv{\'e}n waves propagating along the field. The mechanism is based on {\it swing wave-wave
interaction} [T.V. Zaqarashvili, Astrophys. J. Lett. {\bf 552}, 107 (2001)]. The standing fast magnetosonic waves cause a periodical variation
in the Alfv{\'e}n speed, with the amplitude of an Alfv{\'e}n wave being governed by Mathieu's equation.
Consequently, sub-harmonics of Alfv{\'e}n waves
with a frequency half that of magnetosonic waves grow exponentially in time. It is suggested that the energy of nonelectromagnetic
forces, which are able to support the magnetosonic oscillations, may be transmitted into the energy of purely magnetic
oscillations. Possible astrophysical applications of the mechanism are briefly discussed.
\end{abstract}

\pacs{52.35.Mw}

\maketitle

\section{\label{sec:level1}Introduction}

Many observed phenomena can be associated with wave-like motions, increasing  interest in the study of wave dynamics. Linear perturbation theory considers an
arbitrary disturbance as a superposition of independently evolving eigenmodes, thus simplifying the description
of the process. However, interactions between different harmonics as well as between different kind
of waves leads to the appearance of substantially new phenomena.

In the case of large-amplitude acoustic waves, nonlinearity leads to the generation of higher
harmonics which cause steepening of the wave front and consequently the formation of shock waves. Also developments in plasma theory
raise interest in the study of interactions between different waves. It is shown that nonlinear interaction leads to the generation of resonant triplets (or multiplets) in the plasma \cite{gal,sag,ora}. The nonlinear interaction between magnetohydrodynamic (MHD)
waves has been studied in various astrophysical situations \cite{gold,nak1,nak2,nak3}. Additionally, MHD wave coupling due to
inhomogeneity of the medium \cite{ion,rae,hey,hol,poedts} or
a background flow \cite{chag1,chag2,rog}
has also been developed.

Recently, a new kind of interaction between sound and Alfv{\'e}n waves has been discussed by
Zaqarashvili [16]. The physical basis of this interaction is the parametric influence; sound waves cause a periodical variation in the medium's parameters, which affects the velocity
of transversal Alfv{\'e}n waves and leads to
a resonant energy transformation into certain harmonics.
In a high $\beta$ plasma, it is shown that periodical variations of the medium's density, caused by the propagation of sound waves along an applied
magnetic field, results in Alfv{\'e}n waves being governed by Mathieu's equation (here $\beta={8{\pi}p}/{B^2}\gg1$, where $p$ is the plasma pressure and $B$ is the magnetic field). Consequently, harmonics with half the
frequency of sound waves grow exponentially in time. The same phenomenon was developed in the case of standing sound waves
\cite{zaq2}. The process of energy exchange between these different kinds of wave motion is called
{\it swing wave-wave interaction}. This terminology arises from an analogy with a swinging pendulum, as described below.

In this paper we further develop the theory for interactions between fast magnetosonic waves and Alfv{\'e}n waves. For
clarity of presentation we first recall the pendulum analogy and show that under certain conditions the energy of spring oscillations along the pendulum axis is
transformed into
the energy of transversal oscillations of the pendulum, and {\it vice versa}. Following a discussion of the general physics
of swing interaction we go on to consider the example of coupling between fast magnetosonic waves propagating across an applied magnetic field
and Alfv{\'e}n waves propagating along the field. Finally, we briefly describe the applications of the theory to various astrophysical situations.

\section{Swing pendulum}

It is useful to begin with a mechanical analogy of the wave dynamics in a medium (see \cite{sag} in the case of three-wave interaction). Consider a mathematical pendulum with mass $m$ and equilibrium length $L$ (see Fig. 1). Part of the pendulum
length consists of a spring with stiffness constant $\sigma$. There are two kinds of oscillation in this
system: transversal oscillations due to gravity and spring oscillations along the pendulum
axis due to the elasticity of the spring. This is a {\it swing pendulum}.

In equilibrium, gravity is balanced by the stiffness force $T_0$
of the spring so that
\begin{displaymath}
T_0={\sigma}h=mg,
\end{displaymath}
where $g$ is the gravitational acceleration and $h$ is the equilibrium length of the spring (the natural length of the
spring is supposed negligible).
\begin{figure}
\includegraphics[width=7cm]{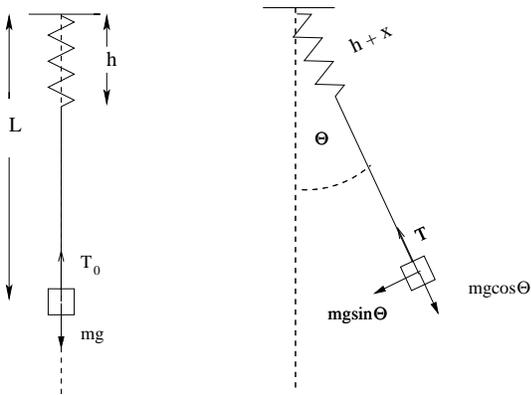}
\caption{\label{fig:epsart}The swing pendulum in equilibrium (left) and in oscillation (right).}
\end{figure}
For displacement $x$ along the pendulum axis, the stiffness force becomes
\begin{displaymath}
T = {\sigma}(h + x)= mg + {\sigma}x.
\end{displaymath}
Newton's second law applied along the pendulum axis, when the pendulum makes an angle ${\Theta}$ with the vertical (see Fig.1),
gives the equation of motion (the centrifugal force due to the transversal oscillation is neglected)
\begin{displaymath}
{\ddot x} + {{\sigma}\over m}x= g({\cos}{\Theta} -1).
\end{displaymath}

Due to the oscillation of the spring along the axis, the pendulum length is
a function of time and the equation of transversal motions of the pendulum under gravity is
\begin{displaymath}
{\ddot {\Theta}} + {{g}\over {L+x}}{\sin}{\Theta}=0.
\end{displaymath}
For clarity of presentation a term $2{\dot x}{\dot {\Theta}}/(L + x)$ is here neglected; it does not affect the physical
nature of the phenomenon (for general consideration, see \cite{mag,zaq3}).
So we have two different oscillations of the pendulum, which are coupled, and each oscillation
influences the other. Considering small amplitude oscillations, we find two coupled equations
governing the dynamics of the pendulum:
\begin{equation}
{\ddot x} + {\omega_1^2}x= -{1\over 2}g{\Theta^2},
\end{equation}
\begin{equation}
{\ddot {\Theta}} + {\omega_2^2}(1 - {{x}\over L}){\Theta}=0,
\end{equation}
where ${\omega_1}={\sqrt {g/h}}$ and ${\omega_2}={\sqrt {g/L}}$ are the fundamental frequencies of the system.

From equations (1) and (2) we can see that longitudinal oscillations of the pendulum causes
a periodical variation of the pendulum length. In certain conditions this can lead
to the well known parametric amplification of transversal oscillations. When $x$ is a periodical function of time, then
equation (2) becames Mathieu's equation and it has a resonant solution when
\begin{equation}
{\omega_2}= {{1}\over 2}\omega_1,
\end{equation}
corresponding to $L=4h$.

Under these conditions, initial spring oscillations, $x$, along the pendulum axis
can amplify small transversal perturbations, $\Theta$ (see equation (2)).
On the other hand, transversal
oscillations may be considered as an external periodic force (see equation (1)) which causes the damping and consequent
amplification of longitudinal oscillations.
So, in the absence of dissipation, there is a subsequent energy exchange between different oscillations in the system.
But if some kind of external force supports the spring oscillations then they can amplify the transversal oscillations until
nonlinear effects became significant.

\section{Swing wave-wave interaction}

The generalisation of the above analogy to waves in a medium leads to interesting phenomena.
Spring oscillations do work against gravity and cause periodical variations of the parameter (pendulum length $L$)
of transversal oscillations.
As a result of this work, the energy of spring oscillations transforms into the energy of transversal oscillations.
So we may expect a similar process in a medium when one kind of waves cause a periodical variation of
another wave parameters.

There are three main forces in the equation of motion for an ideal conductive fluid: the pressure gradient
$-{\nabla}p$, gravity ${\rho}{\nabla}{\phi}$, and the Lorentz force ${\bf j}{\times}{\bf B}$. Here $p$ and ${\rho}$ denote the plasma pressure and density, $\phi$ is the gravitational potential and $\bf j$ is the current in a magnetic field ${\bf B}$. Each of these forces represents the
restoring force against the fluid inertia and
thus leads to the generation of different kinds of wave motion. Of these forces, only the Lorentz force does not include the density (in the pressure gradient
the density arises from the equation of state). This fact leads to the appearance of the density in the expression for the magnetic speed, the
Alfv{\'e}n
speed $V_A=B/{\sqrt{4\pi\rho}}$, which describes the propagation of magnetic waves and depends on the medium density $\rho$. For a similar reason, the frequency of pendulum oscillations does not depend on the pendulum mass (because the gravitational force depends on it), while the frequency of spring oscillation does depend on the mass (because the stiffness force does not depend on it). On the
other hand, compressible waves cause density variations in the medium and therefore they may
affect the propagation properties of magnetic waves. This suggests a coupling between
longitudinal, compressible waves (leading to density perturbations) and transversal magnetic waves propagating with a velocity which depends on the density. The later can be associated with Alfv{\'e}n waves which are transversal and represent the purely electromagnetic properties of the medium. The compressible waves cause a
periodical variation of the density and so of the Alfv{\'e}n speed, and may lead to the effective energy transmission into certain harmonics of
Alfv{\'e}n waves. The swing coupling between sound and Alfv{\'e}n waves propagating along an applied magnetic field [16-17] is a good example of this phenomenon.
On the other hand, magnetosonic waves propagating at an angle to the magnetic field also cause periodical
variations of the Alfv{\'e}n speed and
may lead to similar phenomena. It is worth noticing that, contrary to the Alfv{\'e}n waves, the magnetosonic waves can be easily
excited in a medium by any force (even of non electromagnetic origin). Therefore, the coupling between magnetosonic and Alfv{\'e}n
waves allows the transmission of energy into purely transversal magnetic oscillations through compressible magnetosonic oscillations.

To show the mathematical formalism of swing wave interaction we consider the case of magnetosonic wave propagation across the applied magnetic field. In this case we have fast magnetosonic waves.
For simplicity we consider a rectangular geometry, which then can be generalised to cylindrical and spherical symmetries.

\subsection{Coupling between fast magnetosonic and Alfv{\'e}n waves}

Consider motions of a homogeneous medium, with zero viscosity and infinite conductivity, as described by the ideal MHD
equations:
\begin{equation}
{{{\partial \bf B}}\over {\partial t}} + ({\bf u}{\cdot}{\nabla}){\bf B}=
({\bf B}{\cdot}{\nabla}){\bf u} - {\bf B}{\nabla}{\cdot}{\bf u},\,\,\,\,\,{\nabla}{\cdot}{\bf B}=0,
\end{equation}
\begin{equation}
{\rho}{{{\partial \bf u}}\over {\partial t}} + {\rho}({\bf u}{\cdot}{\nabla})
{\bf u} = - {\bf {\nabla}}\left[p + {{B^2}\over
{8{\pi}}}\right ] + {{({{\bf B}{\cdot}{\nabla}}){\bf B}}\over
{4{\pi}}},
\end{equation}
\begin{equation}
{{{\partial {\rho}}}\over {\partial t}} + ({\bf u}{\cdot}{\nabla}){\rho}
 + {\rho}{\nabla}{\cdot}{\bf u}=0,
\end{equation}
where ${\bf u}$ is the fluid velocity. We consider adiabatic processes,
so the pressure $p$ and density $\rho$ are connected by the relation
\begin{equation}
p=p_0\left ({{\rho}\over {\rho_0}}\right )^{\gamma},
\end{equation}
where $p_0$ and $\rho_0$ are the unperturbed uniform pressure and density and $\gamma$ is the ratio of specific heats. We neglect gravity, though
it may be of importance under some astrophysical conditions.
\begin{figure}
\includegraphics[width=7cm]{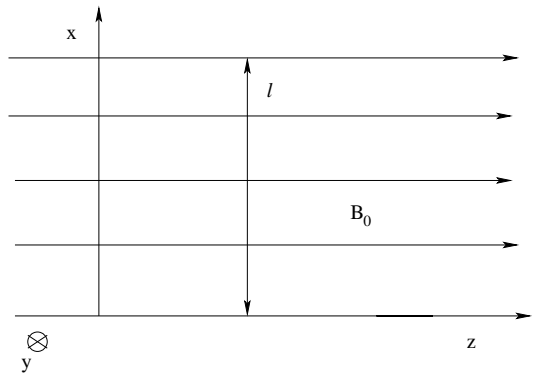}
\caption{\label{fig:epsart}The unperturbed magnetic field $B_0$ is directed along the $z$ axis.The system is bounded in the $x$ direction ($l$ is the size of the system). Standing fast magnetosonic waves are polarized in the $x$ direction, while Alfv{\'e}n waves are polarized along the $y$ axis and propagate along the $z$ axis.}
\end{figure}

Linear analysis of equations (4)-(7) show the existence of three kinds of MHD waves:
Alfv{\'e}n and magnetosonic (fast and slow) waves. The difference between these
waves is that the restoring force of
Alfv{\'e}n waves is the tension of magnetic field lines, $({\bf B}{\cdot}{\nabla}){\bf B}/{4\pi}$, acting alone, while
the restoring force of magnetosonic waves is mainly the gradient of ordinary and magnetic
pressures, $- {\bf {\nabla}}\left[p +
{{B^2}/{8{\pi}}}\right ]$. The various waves can be distinguished by their different speeds and polarizations.
The linear evolution of the waves in a homogeneous medium is governed by the usual linear wave equations.

Consider a uniform, unperturbed, magnetic field ${\bf B_0}=(0,0,B_0)$
directed along the $z$-axis,
and the case
of magnetosonic wave propagation across the field in the $x$ direction (see Fig. 2).
Then there are only fast magnetosonic waves (the slow wave is absent) which in the linear approximation is described by the equations:
\begin{equation}
{{{\partial }{b_z}}\over {{\partial}t}}=
- B_0{{\partial u_x}\over {\partial x}},
\end{equation}
\begin{equation}
{\rho_0}{{{\partial u_x}}\over {\partial t}}= - {{\partial}\over {\partial x}}\left [c_s^2{\rho} +
{{B_0b_z}\over {4{\pi}}}\right ],
\end{equation}
\begin{equation}
{{\partial {\rho}}\over {\partial t}} = - {\rho_0}{{\partial u_x}\over {\partial x}},
\end{equation}
where $b_z$ and $u_x$ are the perturbations
of magnetic field and velocity, respectively, and $c_s=\sqrt {{\gamma}p_0/{\rho_0}}$ is the sound speed. Here and afterwards $\rho$ denotes the perturbation of density (in equations (4)-(5) $\rho$ was the total density).
The wave equation for linear fast magnetosonic waves then follows,
\begin{equation}
{{\partial^2 u_x}\over {\partial t^2}} - V_f^2{{\partial^2 u_x}\over {\partial x^2}}=0,
\end{equation}
where $V_f=\sqrt{c_s^2 + V_A^2}$
is the phase velocity of fast waves and $V_A=\sqrt{B^2_0/4\pi{\rho_0}}$ is the Alfv{\'e}n speed.

The solution of the wave equation can be either propagating or standing patterns. The boundedness of the medium leads to the
formation of a discrete spectrum of harmonics which represent the normal modes (eigenmodes) of the system.
We consider the
standing fast magnetosonic waves, which have a straightforward extention to
cylindrical (pulsating magnetic tube) and spherical (pulsating sphere with dipole-like magnetic field) geometries.
The solutions for standing (plane) fast magnetosonic
waves are:
\begin{equation}
\begin{array}{l}
u_x={\alpha}V_f{\sin}({\omega_n}t){\sin}({k_n}x),\\
{\rho}={\alpha}{\rho}_0{\cos}({\omega_n}t){\cos}({k_n}x),\\
{b_z}={\alpha}B_0{\cos}({\omega_n}t){\cos}({k_n}x), \end{array}
\end{equation}
where $k_n={{n\pi}\over {l}}$ ($n=1,2,...$)
is the eigenvalue for a system of size $l$ in the $x$ direction, $\omega_n$ is the corresponding
eigenfrequency, and $\alpha$ is the relative amplitude of the waves. Eigenvalues and eigenfrequencies are related by
the dispersion relation ${{\omega_n}/{k_n}}=V_f$.

It is seen from the expressions (12) that standing
fast magnetosonic waves
cause a local periodical variation in both the density and the magnetic field.  This variation is maximal near the nodes of
the velocity and approaches to zero near the antinodes. The amplitude of the variation is considered to be small $(\alpha{\ll}1)$, and so
does not affect the fast magnetosonic wave itself.

Consider now the influence of the density and the magnetic field variations (12) on Alfv{\'e}n waves, considered to be polarised in the $yz$ plane. Then the velocity
fields of fast magnetosonic and Alfv{\'e}n waves are decoupled. The
linear equations for Alfv{\'e}n waves are:
\begin{equation}
{{{\partial }{b_y}}\over {{\partial}t}}=
B_0{{\partial u_y}\over {\partial z}},
\end{equation}
\begin{equation}
{\rho_0}{{{\partial u_y}}\over {\partial t}}=  {{B_0}\over
{4{\pi}}}{{\partial b_y}\over {\partial z}},
\end{equation}
where $b_y$ and $u_y$ are small perturbations of the magnetic field and the velocity. These equations lead to the wave equation
\begin{equation}
{{\partial^2 b_y}\over {\partial t^2}} - V_A^2{{\partial^2 b_y}\over {\partial z^2}}=0.
\end{equation}

The influence of the fast magnetosonic waves can be expressed by modifying equations (13) and (14), which now became
\begin{equation}
{{{\partial }{b_y}}\over {{\partial}t}}=
(B_0 + b_z){{\partial u_y}\over {\partial z}} - {{\partial u_x}\over {\partial x}}b_y,
\end{equation}
\begin{equation}
({\rho_0} + {\rho}){{{\partial u_y}}\over {\partial t}}=  {{B_0 + b_z}\over
{4{\pi}}}{{\partial b_y}\over {\partial z}}.
\end{equation}
Here we have neglected the advective terms $u_x{\partial b_y}/{\partial x}$ and
$({\rho_0} + {\rho})u_x{\partial u_y}/{\partial x}$ for several reasons.
At the initial stage, the perturbations $b_y$ and $u_y$ of Alfv{\'e}n waves propagating along the $z$ axis do not depend on the $x$ coordinate; each magnetic surface across $x$ evolves independently. The $x$ dependence arises due to the action of the fast magnetosonic waves, and so the neglected terms are second order in $\alpha^2$. Moreover we can consider the Alfv{\'e}n waves at the velocity node of standing fast magnetosonic waves, where these terms are zero.
In principle, the coordinate $x$ stands as a parameter in equations (16) and (17) of Alfv{\'e}n waves.

Equations (16) and (17) lead to the Hill type second order differential equation
\begin{displaymath}
{{\partial^2 b_y}\over {\partial t^2}} - {{(2B_0 + b_z){\dot b_z}}\over {B_0(B_0 + b_z)}}
{{\partial b_y}\over {\partial t}} - {{(B_0 + b_z){\ddot b_z} - {\dot b_z^2}}\over {B_0(B_0 + b_z)}}b_y -
\end{displaymath}
\begin{equation}
- {{(B_0 + b_z)^2}\over {4\pi({\rho_0} + {\rho})}}{{{\partial^2 b_y}\over {\partial z^2}}}=0,
\end{equation}
where $\dot b_z$ denotes the time derivative of the perturbing field.
Introducing
\begin{equation}
b_y = h_y(z,t)\exp{\int {{{(2B_0 + b_z){\dot b_z}}\over {2B_0(B_0 + b_z)}}dt}}
\end{equation}
and neglecting terms of order $\alpha^2$ leads to the equation
\begin{equation}
{{\partial^2 h_y}\over {\partial t^2}} - V_A^2\left [1 + {\alpha}{\cos}(k_nx){\cos}(\omega_nt)\right ]
{{\partial^2 h_y}\over {\partial z^2}}=0.
\end{equation}
Comparing equations (20) and (15) we can see that the influence of standing fast magnetosonic waves is
expressed through a periodical variation of the Alfv{\'e}n speed.

Performing a Fourier transform of $h_y$ with $h_y=\int{{\hat h_y(k_z,t)}e^{ik_zz}dk_z}$,
equation (20) leads to Mathieu's equation \cite{abr}
 \begin{equation}
{{\partial^2 {\hat h_y}}\over {\partial t^2}} + \left [V_A^2k_z^2 + {\delta}{\cos}(\omega_nt)\right ]{\hat h_y}=0,
\end{equation}
where
\begin{equation}
{\delta}={\alpha}V_A^2k_z^2{\cos}(k_nx),
\end{equation}
with $x$ playing the role of a parameter.
Equation (21) has main resonant solution if
\begin{equation}
{\omega_A}={{B_0k_z}\over {\sqrt{4\pi{\rho_0}}}}={{\omega_n}\over 2}
\end{equation}
and it can be expressed as
\begin{equation}
{\hat h_y}=h_0e^{{{\left |{\delta}\right |}\over {2\omega_n}}t}\left
   [{\cos}{{\omega_n}\over 2}t - {\sin}{{\omega_n}\over 2}t \right ],
\end{equation}
where $h_0=h(0)$. The solution has a resonant character within the frequency interval
\begin{equation}
{\left |{\omega_A} - {{\omega_n}\over 2}
\right |}<{\left |{{\delta}\over {\omega_n}} \right |}.
\end{equation}

Equation (24) shows that the harmonics of Alfv{\'e}n waves with half the frequency of fast magnetosonic waves grow
exponentially in time.
The growth rate of Alfv{\'e}n waves is maximal at the velocity nodes of fast magnetosonic
waves and tends to zero at the antinodes (see equations (12) and (22)). The amplitude of the magnetic field component in Alfv{\'e}n waves
depends on the $x$ coordinate, i.e. there is the periodical magnetic pressure gradient along this direction. Energy
conservation implies that this gradient leads to the damping of initial fast magnetosonic waves, i.e. the energy transformed into Alfv{\'e}n waves is extracted from fast magnetosonic waves. To show this, we
consider the backreaction of amplified Alfv{\'e}n waves on the initial fast magnetosonic waves.

The dependence of $b_y$ on the $x$ coordinate leads to an additional term in the equation of motion (9) for fast waves,
\begin{equation}
{\rho_0}{{{\partial u_x}}\over {\partial t}}= - {{\partial}\over {\partial x}}\left [c_s^2{\rho} +
{{B_0b_z}\over {4{\pi}}}\right ] - {{\partial}\over {\partial x}}\left [{{b_y^2}\over {8{\pi}}}\right ].
\end{equation}
Therefore the wave equation (11) now becomes
\begin{equation}
{{\partial^2 u_x}\over {\partial t^2}} - V_f^2{{\partial^2 u_x}\over {\partial x^2}}= -
{{\partial^2}\over {{\partial t}{\partial x}}}\left [{{b_y^2}\over {8{\pi}}}\right ].
\end{equation}
The additional term has the frequency of the initial fast magnetosonic waves $\omega_n$ (within the order of $\alpha^2$)
and can be considered as the external, periodic force.
At the initial stage it can be neglected as of second order of smallness. However, it becomes
significant because of the exponential growth of amplitudes (see equation (24)). It oscillates out of phase with respect to the initial fast waves (12), thus leading to their damping (as expected from physical considerations).

Note that equations (1) and (2) describing the pendulum oscillations are  similar to
equations (20) and (27) describing Alfv{\'e}n waves and fast magnetosonic waves;
the longitudinal
oscillations of the pendulum correspond to the fast magnetosonic waves and
the transversal oscillations correspond to the Alfv{\'e}n waves.

Swing coupling between fast magnetosonic and Alfv{\'e}n waves may be generalised from rectangular geometry to
other symmetries, though a detailed description is beyond the scope of this paper.

\section{Discussion}

The suggested mechanism of energy transformation from fast magnetosonic into Alfv{\'e}n waves has important consequences.
It can be noted that the Alfv{\'e}n waves hardly undergo either excitation or damping processes, while
magnetosonic waves can be easily excited by external, even nonelectromagnetic, forces. Then swing interaction leads to the intriguing but natural
suggestion that the energy of the nonelectromagnetic force which supports the magnetosonic waves in the system can be transmitted
into the energy of purely magnetic incompressible oscillations.
This result has many astrophysical applications.
We briefly describe several of them.

\subsection{Swing absorption}

Resonant interaction between MHD waves, due to the inhomogeneity of the medium, was proposed by Ionson [8]. It arises
where the frequency of an incoming wave matches the local frequency of the medium. Then a resonant energy
transformation may take place, known as resonant absorption.

Similar phenomenon may arise also due to swing wave interaction. In this case fast magnetosonic waves can transform their energy into Alfv{\'e}n waves,
even in a homogeneous medium. For given medium parameters (magnetic field, density) the energy of fast waves may be {{\lq}absorbed{\rq}} by  harmonics with wavelengths satisfying the resonant condition (23). Consequently, fast magnetosonic waves can transmit their energy into Alfv{\'e}n waves in any spatial distribution of density or magnetic field. The process can be called  {\it swing
absorption}. The particular point of {\it swing absorption} is that energy {\lq}absorption{\rq} occurs through
the harmonics with half the frequency of incoming waves (see equation (23)). The process may be of importance in the Earth's magnetosphere and in the solar atmosphere.

\subsection{Torsional Alfv{\'e}n waves in solar coronal loops}

Swing wave interaction may play an important role in the excitation of torsional Alfv{\'e}n waves in solar coronal loops. It may be suggested that any external action on the magnetic tube, anchored in the highly dynamical photosphere, causes a radial pulsation at the fundamental frequency, like a tuning fork (see also \cite{rob}). For a tube of radius $r_0$ the fundamental frequency of pulsation will be of order ${V_f}/{r_0}$,
where $V_f$ is the phase velocity of fast magnetosonic waves at the photospheric level. If we consider the Alfv{\'e}n and sound speeds to be of order $\sim 10$ kms$^{-1}$ and the radius of order $\sim 10^{2}$ km, then the period of fundamental mode of pulsation will be a few tens of seconds.

Radial pulsations of the tube may lead to the resonant (exponential) amplification of torsional Alfv{\'e}n waves with half the frequency of the pulsations. These high frequency torsional Alfv{\'e}n waves can propagate upward and carry energy from the photosphere into the magnetically controlled corona or they may be damped in chromospheric regions leading to the heating of the chromospheric magnetic network.

\subsection{Coupling between stellar pulsations and torsional oscillations}

Swing wave interaction may be of importance in stellar interiors. A radial pulsation of a spherically symmetric star with dipole-like magnetic
field may lead to the amplification of torsional oscillations.
There are a number of energy sources which can support pulsations: radiation, nuclear reactions, tidal forces in binary stars, convection, etc. (e.g., \cite{cox}). Then the transformation of pulsational energy into torsional oscillations may lead to new sources for stellar magnetic activity.

\section{Concluding remarks}

The swing wave-wave interaction \cite{zaq1} is developed here in the case of fast magnetosonic waves propagating across a magnetic field and Alfv{\'e}n waves propagating along the field. In the case of oblique propagation, slow magnetosonic waves also exist and they may transmit their energy into Alfv{\'e}n waves. In some cases the coupling between slow magnetosonic and Alfv{\'e}n waves may be of importance. Also, the coupling in the case of different geometries (cylindrical, spherical) may be important in astrophysical situations. The most important result of swing wave interaction is that it reveals a new energy channel for Alfv{\'e}n waves, permitting the transformation of energy of nonelectromagnetic origin into the energy of electromagnetic oscillations.

\begin{acknowledgements}
The work was partially supported by INTAS grant N 97-31931 and
NATO Collaborative Linkage Grant No. PST.CLG.976557. T.Z. would
like to thank the Royal Society for financial help.
\end{acknowledgements}

\end{document}